\documentclass[pre,twocolumn,amsmath,amssymb,floatfix]{revtex4-1}

\usepackage{graphicx}
\usepackage{dcolumn} 
\usepackage{bm}      
\usepackage{color}

\begin{document}

\title{Quantification of the strength of inertial waves in a rotating
    turbulent flow}
\author{P.~Clark di Leoni$^1$, P.J.~Cobelli$^1$, P.D.~Mininni$^1$,
    P.~Dmitruk$^1$, and W.H.~Matthaeus$^2$}
\affiliation{$^1$ Departamento de F\'\i sica, Facultad de Ciencias 
Exactas y Naturales, Universidad de Buenos Aires and IFIBA, CONICET, 
Ciudad Universitaria, 1428 Buenos Aires, Argentina. \\
             $^2$ Bartol Research Institute and Department of Physics and 
Astronomy, University of Delaware, Newark, Delaware, U.S.A.}
\date{\today}

\begin{abstract}
We quantify the strength of the waves and their impact on the energy 
cascade in rotating turbulence by studying the wave number and 
frequency energy spectrum, and the time correlation functions of
individual Fourier modes in numerical simulations in three
  dimensions in periodic boxes. From the spectrum, 
we find that a significant fraction of the energy is concentrated in
modes with wave frequency $\omega \approx 0$, even when the external 
forcing injects no energy directly into these modes. However, for modes 
for which the period of the inertial waves $\tau_\omega$ is faster than 
the turnover time $\tau_\textrm{NL}$, a significant fraction of the
remaining energy is concentrated in the modes that satisfy the 
dispersion relation of the waves. No evidence of accumulation of
energy in the modes with $\tau_\omega = \tau_\textrm{NL}$ is observed, 
unlike what critical balance arguments predict. From the time
correlation functions, we find that for modes with 
$\tau_\omega < \tau_\textrm{sw}$ (with $\tau_\textrm{sw}$ the sweeping 
time) the dominant decorrelation time is the wave period, and that
these modes also show a slower modulation on the timescale 
$\tau_\textrm{NL}$ as assumed in wave turbulence theories. The rest 
of the modes are decorrelated with the sweeping time, including the very
energetic modes modes with $\omega \approx 0$.
\end{abstract}
\maketitle

\section{\label{sec:Intro}Introduction}

Restitutive forces in an incompressible fluid give rise to the development of
waves when the fluid is slightly perturbed from its state of rest. That is the
case of inertial waves in rotating fluids, inertial gravity waves in stratified
fluids, or Alfv\'en waves in conducting fluids. However, when the perturbation
is large, the system can develop far from equilibrium dynamics, with the waves
coexisting with eddies in a fully developed turbulent flow. In such a case, and
when the wave period is much faster than the turnover time of the waves, wave
turbulence theories can be used to predict the scaling laws followed by the
system \cite{Cambon04}.

In the case of rotating flows, the presence of background rotation breaks down
isotropy, and a preferred direction arises along the axis of rotation (see,
e.g., \cite{Cambon89,Cambon97,Bellet06}). As a result of a selection of triadic
interaction by resonant waves, energy is preferentially transferred towards
modes in Fourier space in the plane perpendicular to the rotation axis
\cite{Waleffe93}. The transfer of energy, besides becoming anisotropic, is also
slowed down, resulting in a steeper energy spectrum than in the isotropic and
homogeneous case \cite{Waleffe93,Cambon97}.

Generally speaking, wave turbulence theories can be separated into theories 
of weak and of strong turbulence. In the former case, the assumption of 
weak nonlinearities results in decorrelation between the modes being 
governed by linear dispersion, and the equations can be closed to obtain 
exact spectral solutions. For rotating turbulence, the theory of weak
turbulence predicts an axisymmetric energy spectrum 
$e(k_\perp, k_\parallel) \sim k_\parallel^{-1/2} k_\perp^{-5/2}$ \cite{Galtier03}, 
but this theory only applies to a subset of modes dominated by waves. 
Moreover, as energy may be transfered outside this subset of modes at 
finite time, whether the turbulence can remain weak in rotating flows
has been a matter of debate \cite{Chen05}. In the latter case, theories of 
strong turbulence can describe modes with eddy turnover time of the order
of the wave period (although not the modes with zero frequency in the 
waves), and give a more complete description of the flow, but rely on 
phenomenological approximations to obtain energy spectra that are
positive definite \cite{Cambon89,Cambon97,Cambon04}.

Numerical simulations and experiments give results that in some cases are
consistent with some of the predictions of wave turbulence theories
\cite{Muller07,Mininni09,Mininni12}, but the strength and relevance of the waves
is hard to quantify. In this paper we quantify the strength of inertial waves
and their impact on the turbulent dynamics of rotating flows in 
numerical simulations in periodic boxes by two means: (1)
We compute wave number and frequency spectra, in which the dispersion 
relation of the waves can be directly observed (cf.
\cite{Cobelli2009,Cobelli2011,Dmitruk09}). Using these spectra, the amount of
energy in waves and in eddies can be discriminated. (2) We compute time
correlation functions of individual modes in Fourier space, to identify the
relevant decorrelation time depending on the scale.

Besides the analysis presented here to identify the role of the waves
in setting the dominant timescale in a rotating flow, it is important to 
note that a proper understanding of decorrelation times in turbulence 
is relevant for many applications, as well as for other theoretical 
approaches to turbulence. A somewhat different approach to study 
turbulence in the presence of waves is that of Rapid Distortion Theory 
(RDT). In RDT, the presence of a time scale in the fluid which is much 
shorter than the turnover time (or the decorrelation time) of the large 
scale eddies allows certain magnitudes in a turbulent flow to be
computed using linear theory (see, e.g.,
\cite{cambon_linear_1999,Durbin} for reviews, and \cite{Cambon89} 
for the specific case of rotating turbulence).

Time correlation functions and decorrelation times were computed 
before for rotating fluids, with the focus on their relevance to
predict the acoustic emission produced by a turbulent flow, and 
on the effect of flow anisotropy in the decorrelation 
time \cite{Favier10}. For magnetohydrodynamic flows, time 
correlation functions and decorrelation times were recently computed 
in \cite{Servidio11}. In isotropic and homogeneous turbulence, a
proper understanding of the decorrelation time is needed to correctly 
obtain the frequency spectrum from the Kolmogorov spectrum in terms 
of wavenumbers \cite{Tennekes75}. In this latter case, the dominant 
timescale for all modes is the sweeping time, associated with the 
interactions of the small-scale eddies with the large-scale energy 
containing eddies \cite{Chen89,Nelkin90,Sanada92}. Finally, time 
correlation functions are also important in turbulence closure models, 
for the dynamics of Lagrangian particles \cite{Monin}, and for the 
computation of turbulent diffusion of passive scalars (see, e.g., 
\cite{Blackman03}).

\section{Rotating flows}

\subsection{Waves and eddies}

The dynamics of incompressible rotating flows is described by the Navier-Stokes
equations in a rotating frame,
\begin{equation}
    \frac{\partial {\bf u}}{\partial t} = - \mbox{\boldmath $\omega$} \times
        {\bf u} - 2 \mbox{\boldmath $\Omega$} \times {\bf u} 
        - \nabla {\cal P} + \nu \nabla^2 {\bf u} + {\bf F} ,
\label{eq:momentum}
\end{equation}
together with the incompressibility condition
\begin{equation}
    \nabla \cdot {\bf u} =0 .
    \label{eq:incompressible}
\end{equation}
In these equations, ${\bf u}$ is the velocity, 
$\mbox{\boldmath $\omega$} = \nabla \times {\bf u}$ is the vorticity, 
${\cal P}$ is the total pressure (including the centrifugal term, and 
normalized by the uniform fluid mass density), $\Omega$ is the
rotation frequency, the rotation axis is in the $z$ direction with 
$\mbox{\boldmath $\Omega$} = \Omega \hat{z}$, ${\bf F}$ is an
external mechanical force per unit of mass density, and $\nu$ is the 
kinematic viscosity.

Solutions to these equations can be characterized by two dimensionless
parameters, the Reynolds number 
\begin{equation}
    \textrm{Re} = \frac{UL}{\nu} ,
\end{equation}
and the Rossby number
\begin{equation}
    \textrm{Ro} = \frac{U}{2L \Omega} ,
\end{equation}
where $U$ is the r.m.s.~velocity, and $L$ is the forcing scale.

\begin{figure}
    \centering
    \includegraphics[width=8cm]{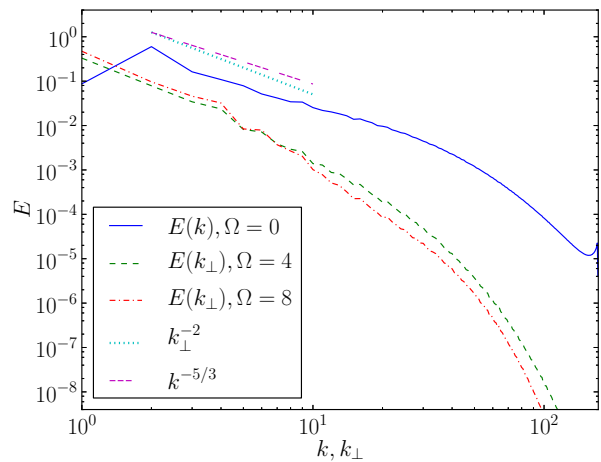} 
    \caption{{\it (Color online)} Isotropic energy spectrum 
        $E(k)$ in the simulation with $\Omega=0$, and reduced 
        perpendicular energy spectra $E(k_\perp)$ in the simulations 
        with $\Omega=4$ and $8$. In all three simulations $\textrm{Re} \approx
        5000$, while $\textrm{Ro} \approx \infty$, $0.03$, and $0.015$ 
        respectively. Kolmogorov and $\sim k_\perp^{-2}$ slopes are shown as a
        reference. The simulation without rotation has a spectrum with a narrow
        range of scales arguably compatible with Kolmogorov scaling and followed
        by a bottleneck and a dissipative range, while the runs with rotation
        display a steeper spectrum.}
    \label{fig:spectrum}
\end{figure}

In the ideal case and in the absence of forcing, the linearized equations have
helical waves ${\bf h}_s$ as solutions, with $s=\pm 1$ corresponding
to the two possible circular polarizations such that 
$i {\bf k} \times {\bf h}_s = s k {\bf h}_s$, and with ${\bf k}$ the
wave vector. These waves correspond to inertial waves with dispersion 
relation $\omega_{\bf k}= s 2 \Omega k_z/k$. The velocity field at 
wave vector ${\bf k}$ can then be decomposed as \cite{Waleffe93}
\begin{equation}
    {\bf u}({\bf k},t) = a_+({\bf k},t) {\bf h}_+ + a_-({\bf k},t) {\bf h}_- .
\end{equation}
In the nonlinear case, a large number of modes are excited (and nonlinearly
coupled) in the velocity field. As a rotating flow can sustain both waves and
eddies, for sufficiently strong rotation it is safe to assume that for a large
number of wave vectors ${\bf k}$ the waves will be faster than the eddies. Then,
in wave turbulence theories the amplitudes $a_s({\bf k},t)$ are further
decomposed into
\begin{equation}
    a_s({\bf k},t) = A_s(T) e^{i \omega_{\bf k} t} ,
    \label{eddiewaves}
\end{equation}
where $e^{i \omega_{\bf k} t}$ is the fast variation at timescale $\tau_\omega
= 2 \pi/\omega_{\bf k}$ associated with the waves, and $A_s(T)$ is a
slowly varying modulation on a timescale $T \sim \textrm{Ro} \, t$ 
associated with the eddies.

Replacing this decomposition in Eq.~\eqref{eq:momentum}, it is obtained that
energy is only transferred between modes with wave vectors ${\bf k}$, ${\bf p}$,
and ${\bf q}$ such that \cite{Cambon89,Waleffe93,Cambon97}
\begin{equation}
    {\bf k} + {\bf p} + {\bf q} = 0,
    \label{eq:triad}
\end{equation}
and
\begin{equation}
    \omega_{\bf k} + \omega_{\bf p} + \omega_{\bf q} = 0.
    \label{eq:resonance}
\end{equation}
The last relation, corresponding to the resonant condition of the waves to have
net transfer of energy when integrated over times longer than the wave period,
is also associated with the development of anisotropies in the flow. Equation
\eqref{eq:resonance} is trivially satisfied for modes with $k_z=0$ (the
so-called 2D or ``slow'' modes, as those modes have wave frequency 
$\omega_{\bf k} = 0$), and Eqs.~\eqref{eq:triad} and \eqref{eq:resonance} drive the
nonlinear coupling to transfer energy preferentially towards modes with small
$k_z$ \cite{Waleffe93}. However, the problem with wave turbulence theories is
that they are not valid for small values of $k_z$, as those modes have eddy
turnover times of the order (or faster) than the wave periods. In fact, in many
theories the predicted energy transfer towards modes with $k_z=0$ vanishes, and
2D modes are then completely decoupled from wave modes \cite{Galtier03}.

\begin{figure}
    \centering
    \includegraphics[width=8cm]{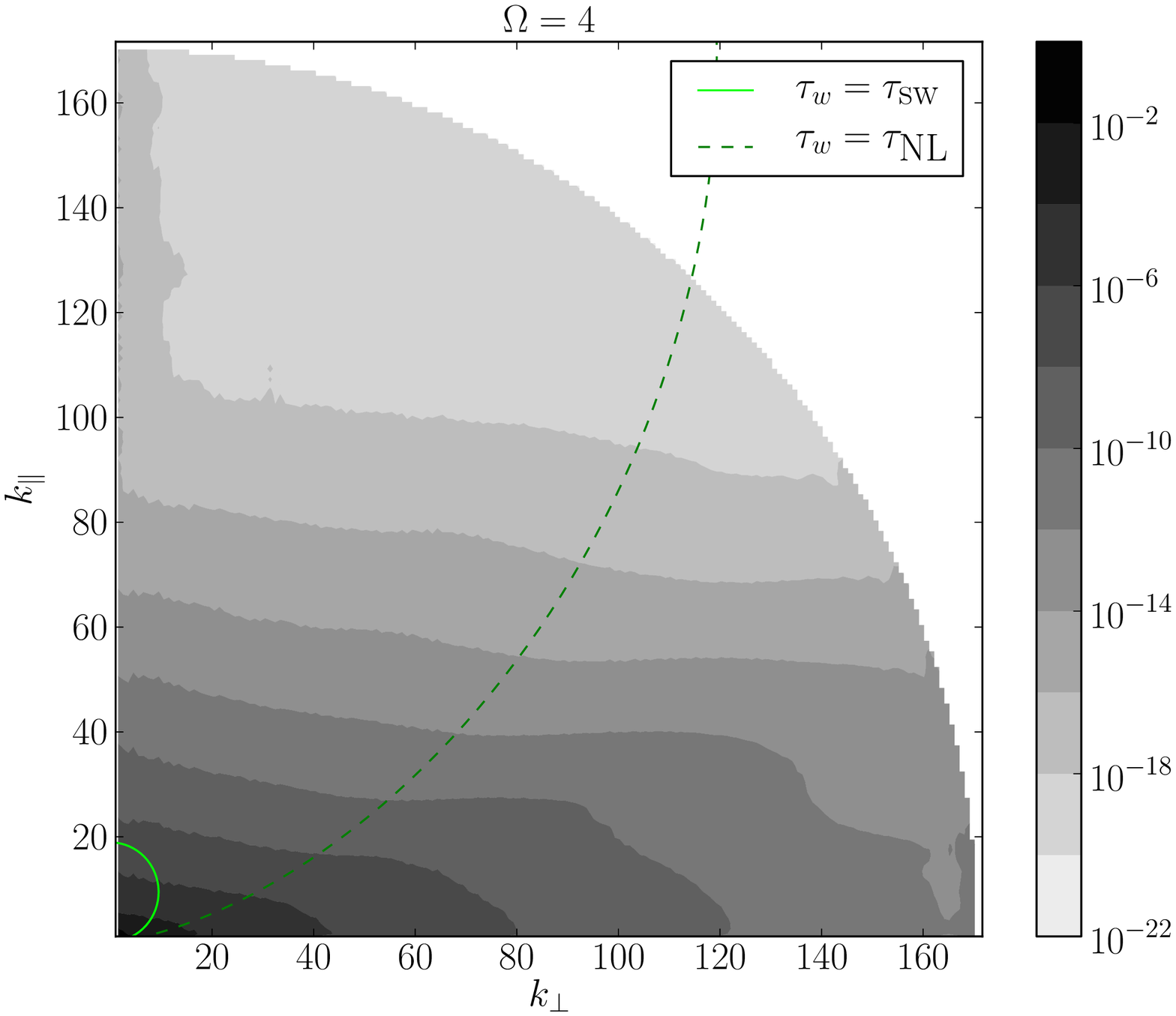} 
    \includegraphics[width=8cm]{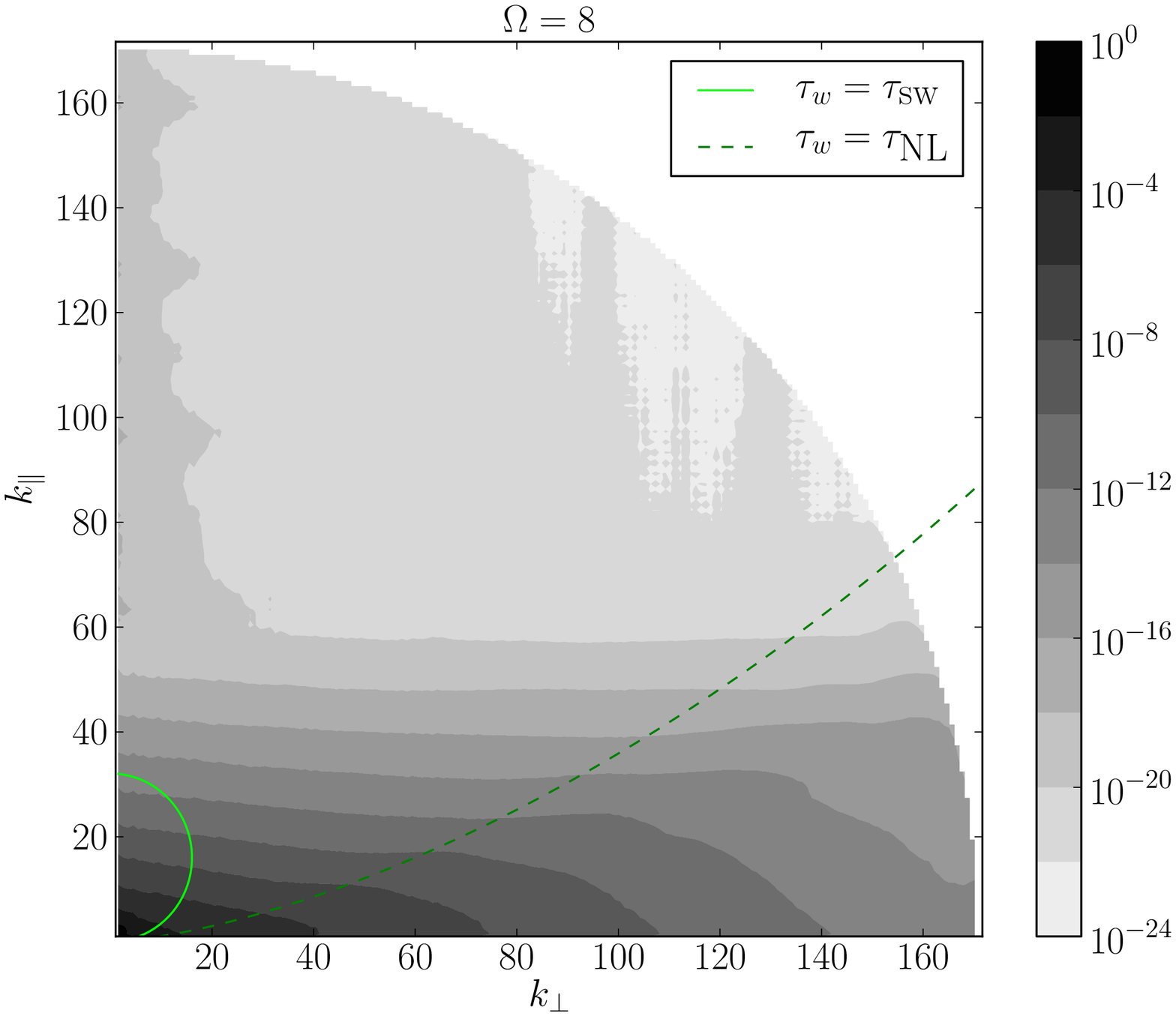} 
    \caption{{\it (Color online)} Isocontours of the axisymmetric
        energy spectrum $e(k_\perp,k_\parallel)$ in the runs with 
        $\Omega=4$ (above) and $8$ (below); dark means larger energy
        density (in logarithmic scale). Lines indicating the modes for
        which the wave time becomes equal to the sweeping time, and to 
        the turnover time, are given as references. It should be noted
        that the energy does not accumulate near the modes with 
        $\tau_\omega = \tau_\textrm{NL}$, unlike what is expected in
        theories dealing with the concept of critical balance 
        \cite{Nazarenko11}.}
    \label{fig:2Dspectrum}
\end{figure}

At this point, it is important to distinguish more precisely between
theories of weak and of strong turbulence. In weak turbulence theory 
it is assumed that rotation is so strong that nonlinear interactions
are weak, such that the linear decorrelation of the waves dominates
over any nonlinear decorrelation time. Such theories predict the
development of anisotropy and the transfer towards modes with smaller 
$k_z$, but the transfer is arrested as the energy reaches modes with
turnover time of the order of the wave period (see, e.g., \cite{Galtier03}). In
theories of strong turbulence (as, e.g., Eddy-Damped Quasi-Normal 
Markovian, or EDQNM, closures) strong nonlinear coupling can be
modeled (although the modes with $k_z=0$ still remain decoupled from 
the other modes), and nonlinear decorrelation times can be dominant
\cite{Cambon89,Cambon97,Bellet06}. However, for the energy to remain 
positive in such closures a damping time must be externally imposed 
(see \cite{Cambon04}). This damping time is often chosen as
\begin{equation}
    \frac{1}{\tau_D} = \sqrt{\sum_i \left(\frac{1}{\tau_i}\right)^2} ,
    \label{eq:EDQNMdamping}
\end{equation}
where $\tau_i$ are the different times in the system (e.g., viscous,
wave, and eddy turnover times), although other empirical combinations
can be used \cite{Cambon89,Cambon97,Bellet06} to improve the 
modeling of spectral anisotropy.

\begin{table}
    \caption{Fraction of the energy contained in the different regions of
        ${\bf k}$-space defined in Fig.~\ref{fig:2Dspectrum}, for the
        two runs with $\Omega \neq 0$. $\Omega$ is the rotation
        frequency, $E(\textrm{2D})/E$ is the ratio of energy in the 2D
        modes to the total energy, $E(\tau_\omega < \tau_\textrm{sw})/E$ is 
        the fraction of energy in the modes with the wave period faster than 
        the sweeping time, and $E(\tau_\omega < \tau_\textrm{NL})/E$ is the
        fraction of the energy in the modes with the wave period faster than 
        the turnover time.}
    \label{table:energies}
    \centering
    \begin{ruledtabular}
    \begin{tabular}{c c c c}
        $\Omega$ & $E(\textrm{2D})/E$ & $E(\tau_\omega < \tau_\textrm{sw})/E$ 
            & $E(\tau_\omega < \tau_\textrm{NL})/E$  \\
        \colrule
        4 & 0.37 & 0.59 & 0.60 \\
        8 & 0.31 & 0.67 & 0.68 \\
    \end{tabular}
    \end{ruledtabular}
\end{table}

\subsection{Wavenumber-frequency spectrum and correlation functions}

Rotating turbulence is often studied using spectra of spatial fluctuations,
either isotropic or anisotropic. However, distinction between eddies and waves
requires spectra also resolved in frequencies, to distinguish the modes
that satisfy the dispersion relation from the rest. There are studies in which 
the presence of inertial waves was explicitly verified in simulations
and experiments \cite{Hopfinger82,Godeferd99,Manders03} (including
observations of inertial waves in the Earth core \cite{Aldridge87}),
although studying their coexistence with turbulent eddies is only
possible with large amounts of data. This can be understood as
computation of spectra resolved in time and in space require storing
data of high resolution simulations (or experiments) with a very short 
cadence in time (at least twice faster than the fastest waves in the
system), and for very long times (at least twice the slowest timescale
in the flow).

In the following we present spectra $E_{ij}({\bf k}, \omega)$ for several
numerical simulations, defined as
\begin{equation}
    E_{ij}({\bf k}, \omega) = 
        \frac{1}{2} \hat{u}_i^*({\bf k}, \omega) \hat{u}_j({\bf k}, \omega),
\end{equation}
where $\hat{u}_i({\bf k}, \omega)$ is the Fourier transform in time and in space
of the $i$-component of the velocity field ${\bf u}({\bf x},t)$, and where the
asterisk denotes complex conjugate. 

\begin{figure*}
    \centering
    \includegraphics[width=7cm]{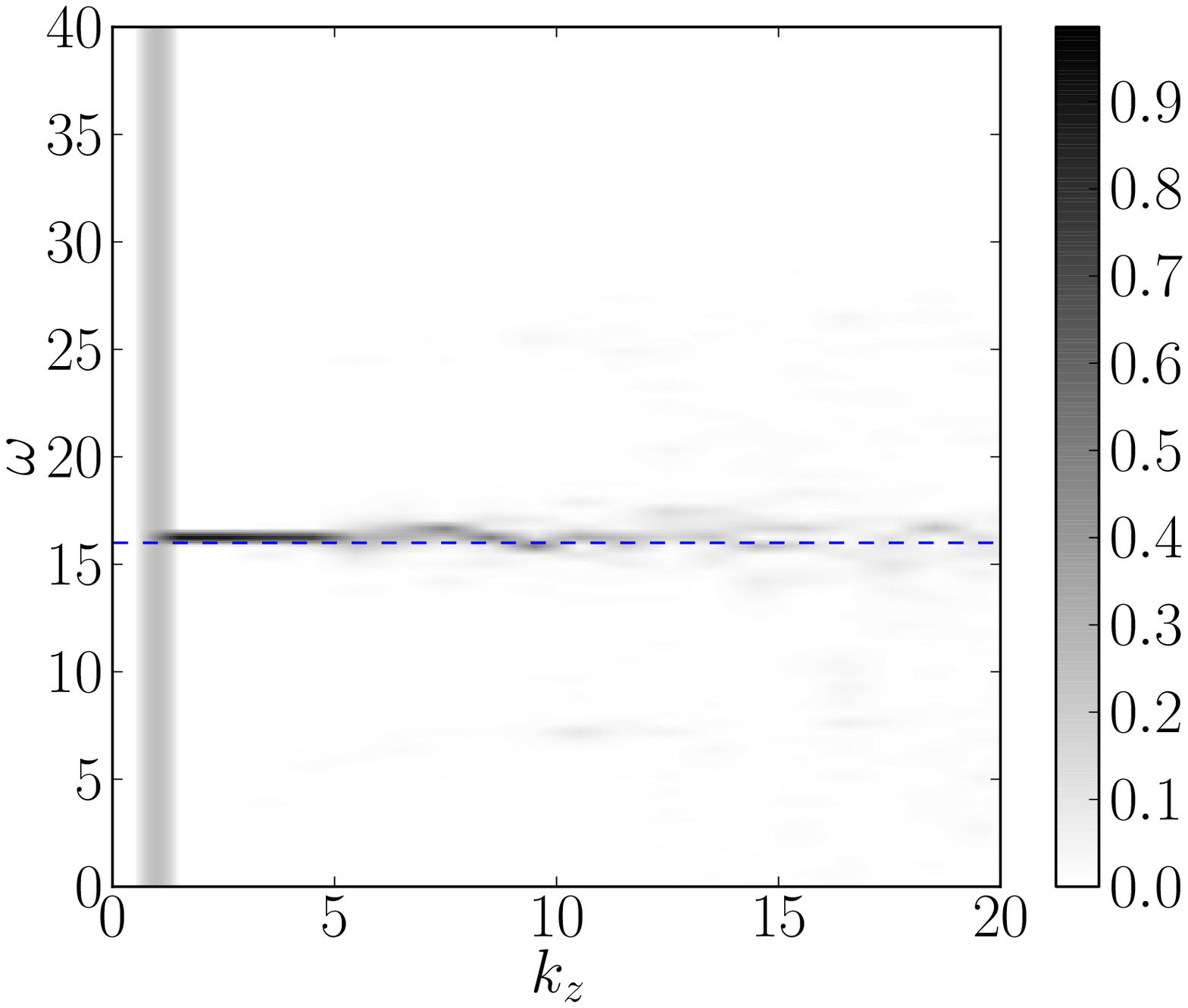}
    \includegraphics[width=7cm]{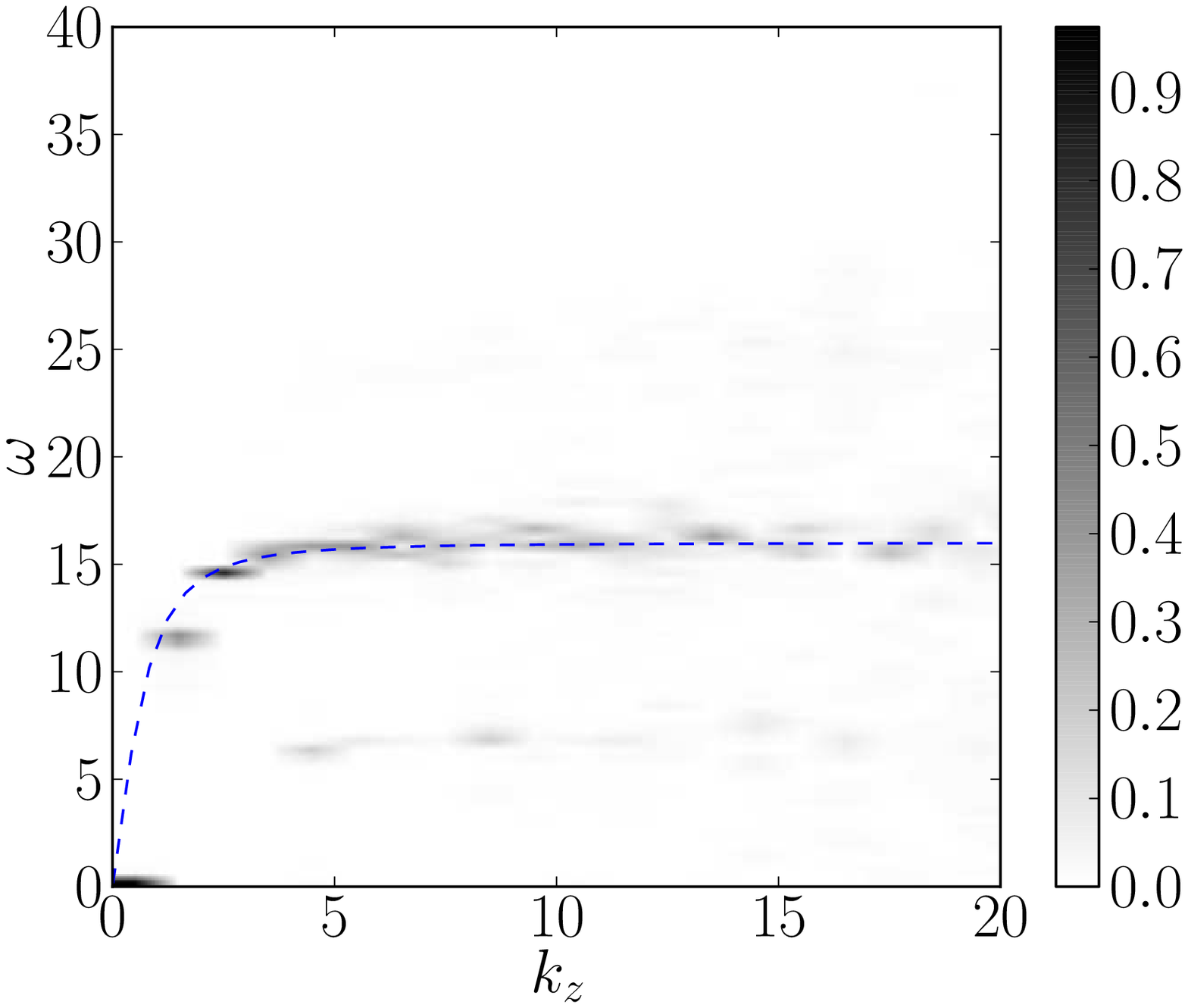} \\
    \includegraphics[width=7cm]{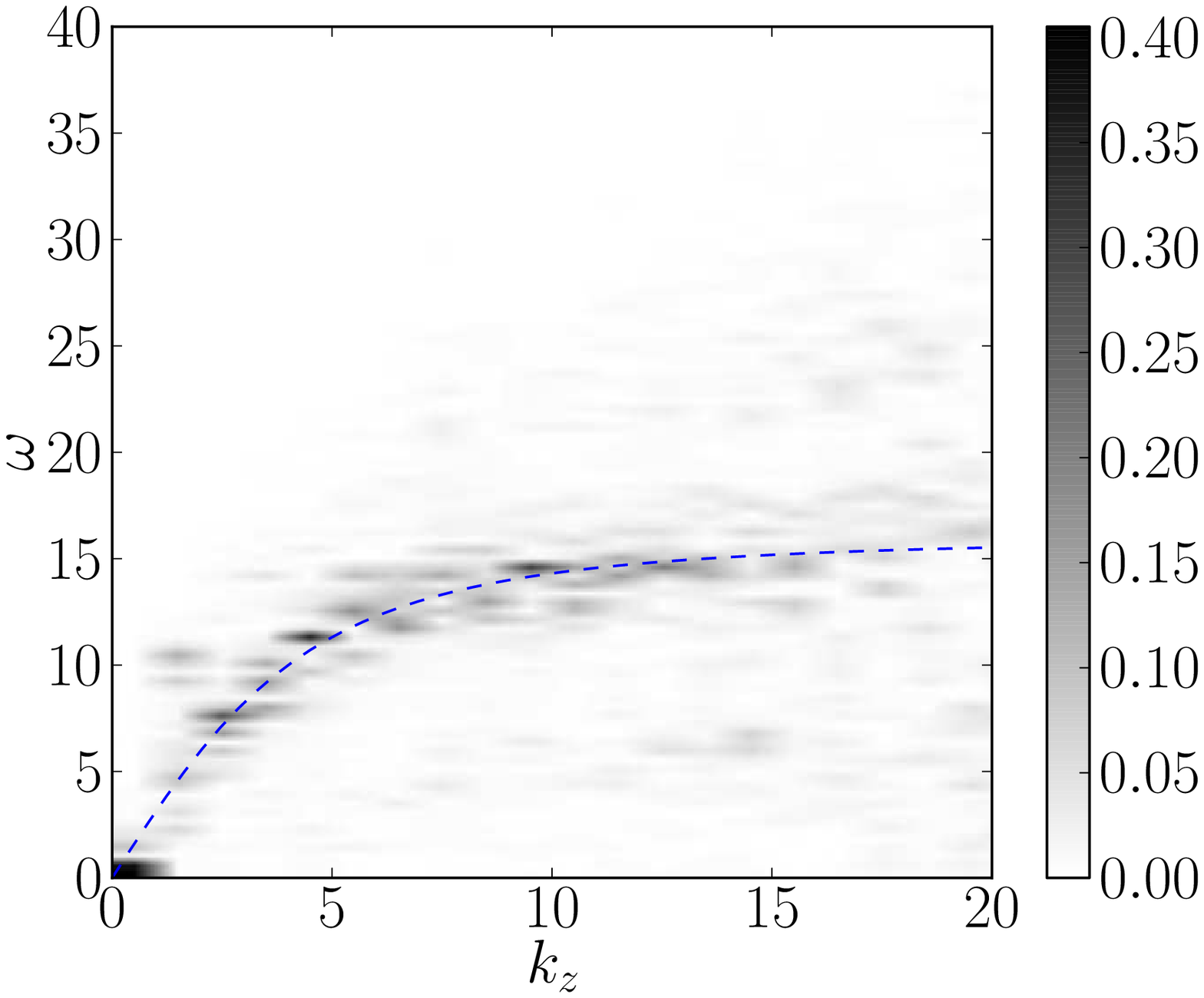}
    \includegraphics[width=7cm]{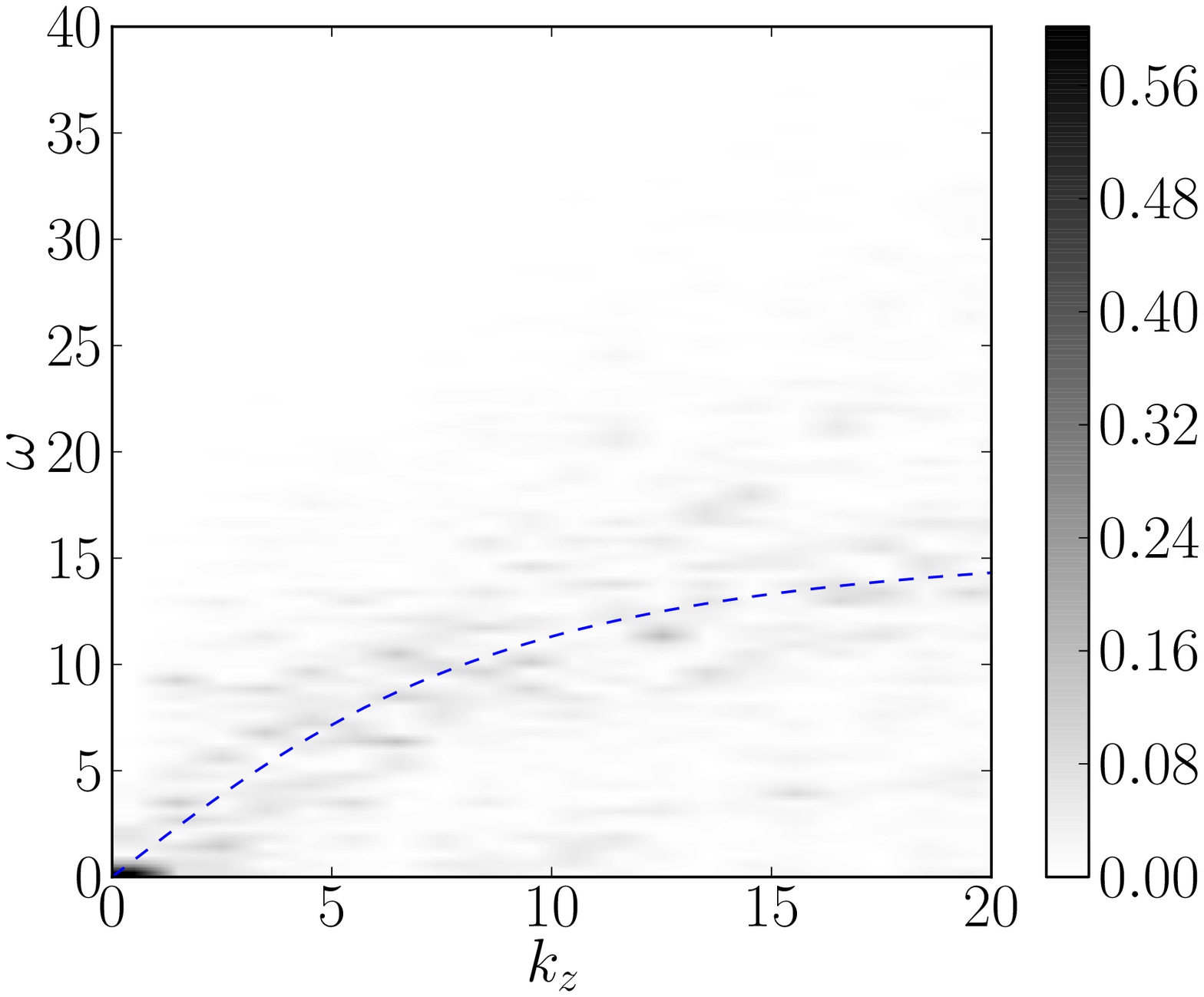}
    \caption{{\it (Color online)} Normalized wave vector and frequency 
        spectrum $E_{11}({\bf k}, \omega)/E_{11}({\bf k})$ for the run with 
        $\Omega=8$. Darker regions indicate larger energy density. The 
        dashed curve indicates the dispersion relation for inertial waves. 
        {\it Top left:} Normalized $E_{11}(k_x=0, k_y=0, k_z, \omega)$.
        {\it Top right:} Normalized $E_{11}(k_x=0, k_y=1, k_z, \omega)$.
        {\it Bottom left:} Normalized $E_{11}(k_x=0, k_y=5, k_z, \omega)$.
        {\it Bottom right:} Normalized $E_{11}(k_x=0, k_y=10,
        k_z,\omega)$. Note from the maximum values in the color 
        bars how the modes close to the dispersion relation
        concentrate most of the energy in the first two cases ($k_y =
        0$ and $k_y = 1$), while as $k_y$ is increased energy becomes
        more spread.
       }
    \label{fig:wspectra}
\end{figure*}

Information on the relevant timescales for each spatial mode, and on their
decorrelation time, can be obtained also from the time correlation function
\begin{equation}
    \Gamma_{ij}({\bf k}, \tau) = 
        \frac{\left< \hat{u}_i^*({\bf k}, t) \hat{u}_j({\bf k}, t+\tau)
            \right>_t}
        {\left< |\hat{u}_i^*({\bf k}, t) \hat{u}_j({\bf k}, t)| \right>_t} ,
\end{equation}
where  $\hat{u}_i({\bf k}, t)$ is the Fourier transform in space of the
$i$-component of the velocity field, the brackets denote time average, and only
the real part is used. If the mode $\hat{u}_i({\bf k}, t)$ is dominated by waves
in a regime that satisfies the hypothesis of weak turbulence theory, then
$\Gamma_{ii}({\bf k}, \tau) \sim \cos(\omega_{\bf k} \tau)$. If nonlinear
effects are important, then the mode with wave vector ${\bf k}$ should be
decorrelated after a time $\tau_D({\bf k})$ following an approximate exponential
decay
\begin{equation}
    \Gamma_{ii}({\bf k}, \tau) \sim e^{-\tau/\tau_D({\bf k})} .
\end{equation}
In the following we will define $\tau_D$ as the time at which the function
$\Gamma$ decays to $1/e$ of its initial value. Note this definition is
arbitrary, and some authors use the half-width of the correlation
function, or a value based on an integral timescale (see, e.g., 
\cite{Favier10,Servidio11})
\begin{equation}
\tau_D({\bf k}) = \int_0^\infty \Gamma_{ii}({\bf k}, \tau) \, \text{d} \tau .
\end{equation}
We verified that no quantitative differences are obtained by using
these other definitions, except for a multiplicative factor of order
one in the values of all decorrelation times.

\begin{figure}[h!]
    \centering
    \includegraphics[width=8cm]{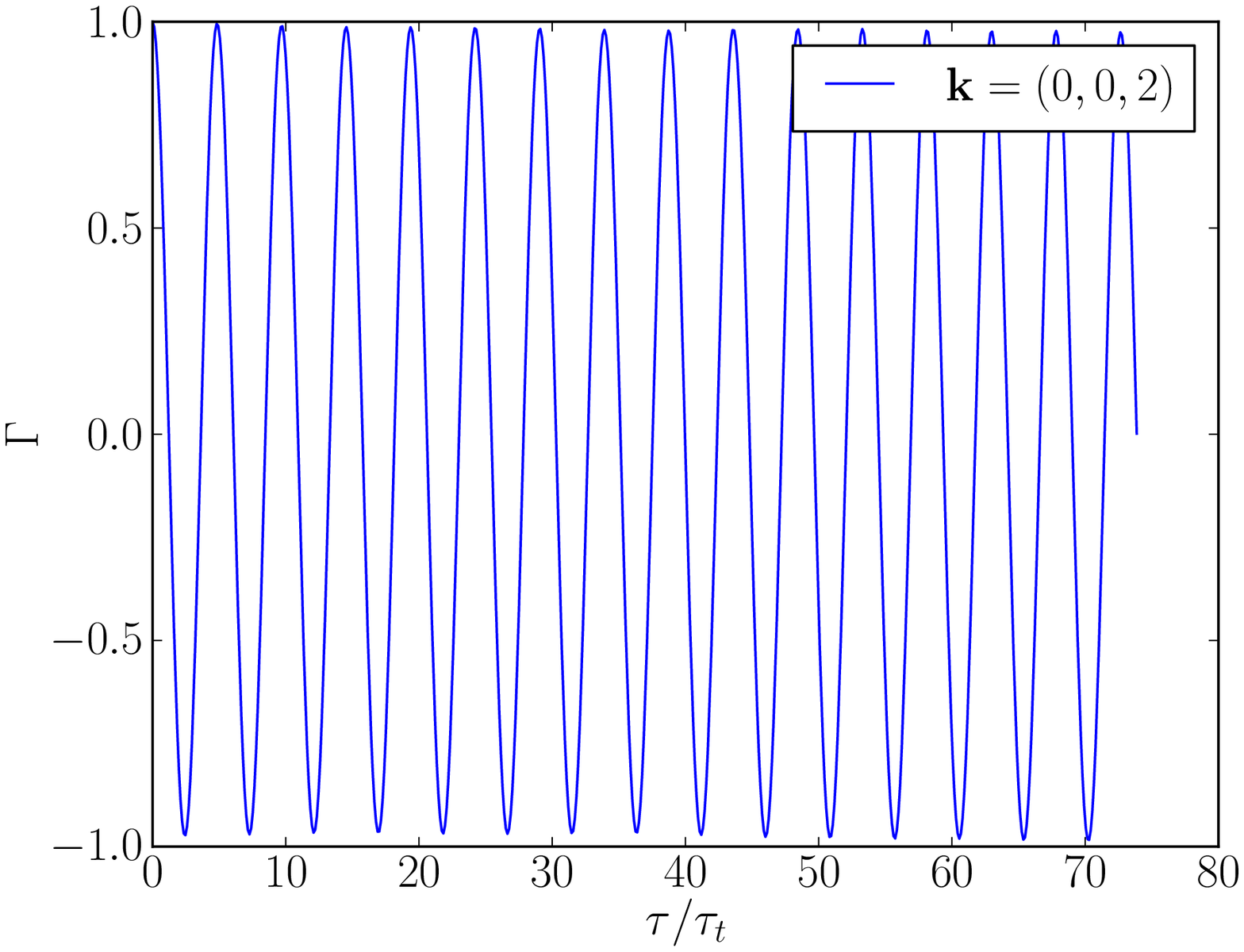} \\
    \includegraphics[width=8cm]{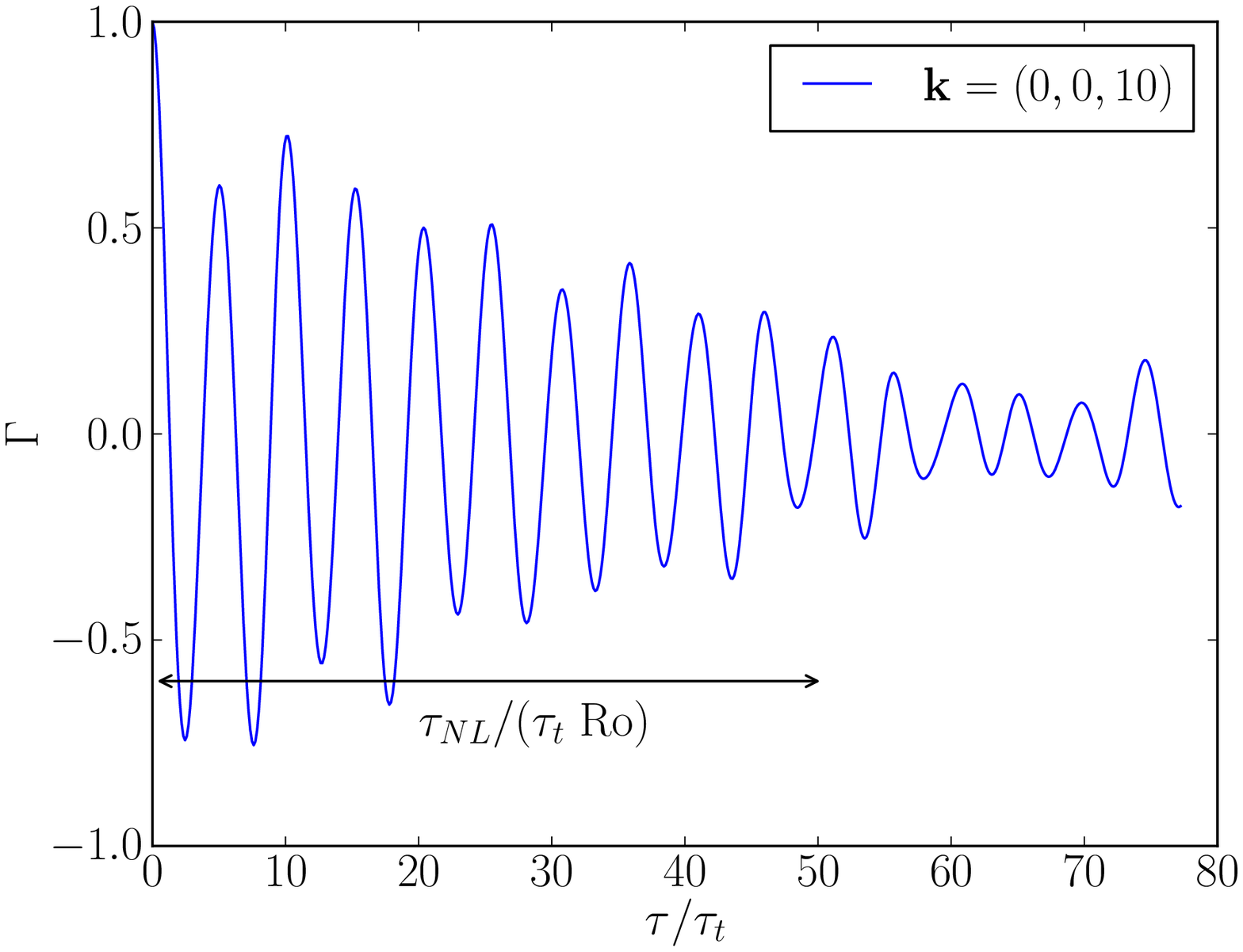} \\
    \includegraphics[width=8cm]{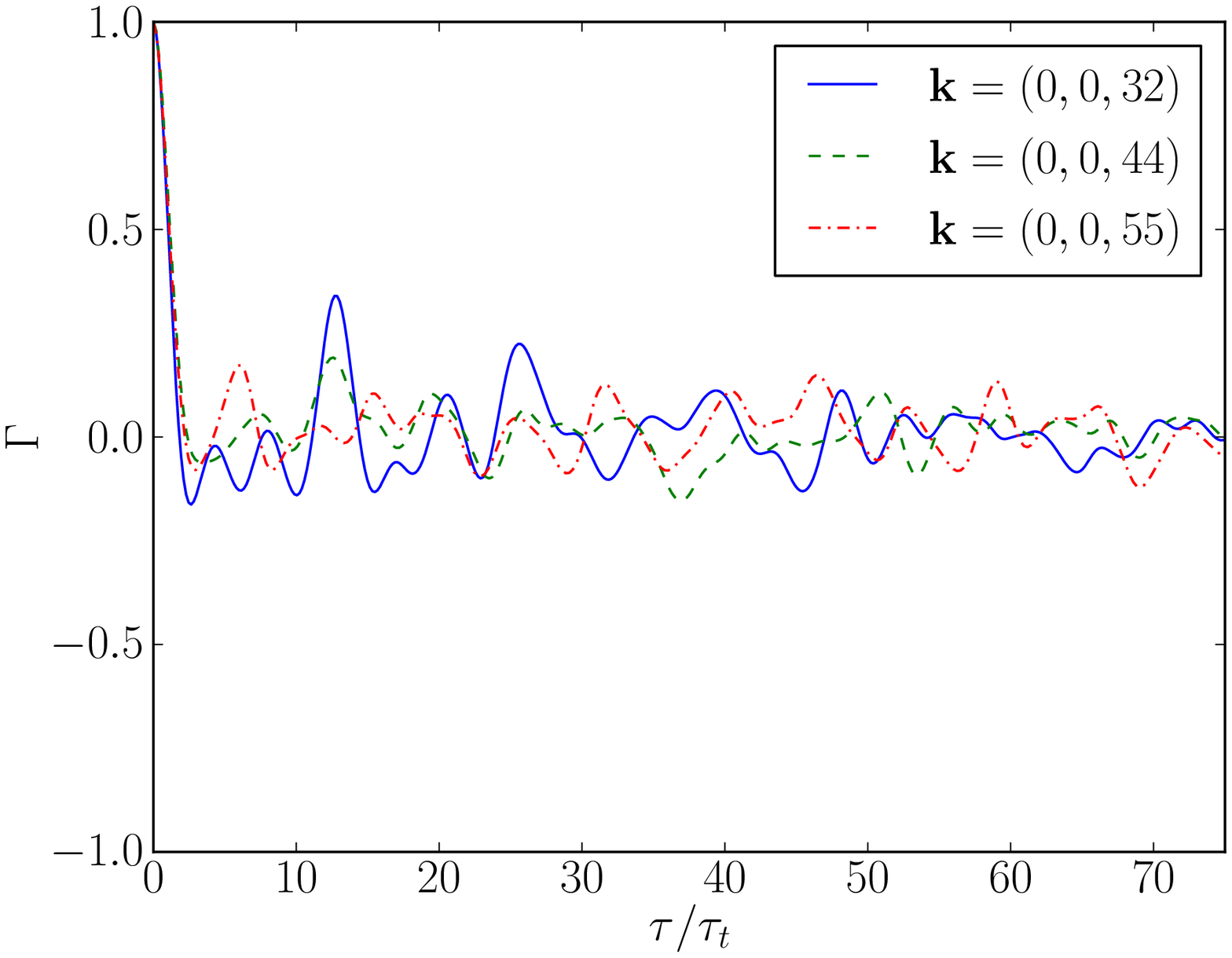}
    \caption{{\it (Color online)} Correlation function $\Gamma_{11} (\tau)$ 
        for different modes, the wave-like behavior is quite evident in 
        the first two panels. Time is normalized such that 
        $\Gamma_{11} (\tau) = 1/e$ at a time of order unity. 
        See text for the definition of $\tau_\textrm{t}$. {\it Top:} A 
        mode with $\tau_\omega \ll \tau_\textrm{sw}$. {\it Middle:} A
        mode with $\tau_\omega < \tau_\textrm{sw}$. Here $\tau_\omega$ 
        still dominates, and the correlation function shows the oscillating 
        behavior expected for a wave-like mode, but with a slow
        decay in its modulation proportional to the nonlinear time 
        (indicated by the arrow). {\it Bottom:} A mode with 
        $\tau_\omega > \tau_\textrm{sw}$, here all wave-like behavior
        is lost.}
    \label{fig:correlation}
\end{figure}

\subsection{Numerical simulations}
\label{numsimuls}

Computation of the functions described above require a significant amount of
storage. As a result, only moderate resolution simulations can be performed. We
performed three simulations using grids of $N^3=512^3$ points, in a
three-dimensional periodic box. 

Equations (\ref{eq:momentum}) and (\ref{eq:incompressible}) were
solved using a parallel pseudospectral method, and evolved in time 
with a second order Runge-Kutta scheme (for more details 
of the code, see \cite{Gomez05,Mininni11}). The simulations were 
dealiased with the $2/3$-rule (see, e.g., \cite{Gomez05}).

The equations are written in dimensionless units. The periodic 
domain has length $\lambda_0 = 2\pi$, resulting in integer wavenumbers and 
in a minimum wave number $k_\textrm{min} = 2\pi/\lambda_0 = 1$. Per virtue 
of the $2/3$-rule, the largest resolved wave number is 
$k_\textrm{max} = N/3$, associated with the smallest resolved 
wavelength $\lambda_\textrm{min} = 2\pi/k_\textrm{max} = 6\pi/N$. With 
this choice, for a characteristic velocity $U_0=1$ and a characteristic
length $L_0=1$, the turnover time is $T_0 = L_0/U_0 =1$, which we use 
as unit of time. $\Omega$ is then measured in units of the inverse of 
time $T_0$.

In previous studies of rotating turbulence in periodic domains, it was found
that if the forcing is applied at intermediate scales (i.e., scales smaller than
the size of the domain), an inverse cascade develops and most of the energy ends
up in the 2D modes \cite{Sen12}. Evidence of this inverse cascade has been also
observed in experiments \cite{Yarom13}. It is unclear for the moment whether
this effect also takes place in homogeneous, unbounded flows, such as those
considered by wave turbulence theories \cite{Cambon04}. As a result,
we forced the system at the largest scales available, to prevent the
inverse cascade from developing. However, this has a caveat: the
finite domain selects a discrete set of inertial waves which are
normal modes of the domain (see, e.g., \cite{Smith05,Bellet06}). As a
result of the discretized wavenumbers, the number of modes that satisfy
the resonance condition \eqref{eq:resonance} depends on the wavenumber, and is
smaller (or zero) for smaller wavenumbers, resulting in only near-resonances
being available \cite{Smith05}. As this effect is aggravated when domains with
non-unity aspect ratio are used, we restricted our study to boxes with aspect
ratio of unity.

As we are also interested in correlation times, to prevent imposing external
correlation times with the forcing we used a coherent forcing (in opposition to a
time-correlated, or delta-correlated in time forcing function). We therefore used
Taylor-Green forcing
\begin{eqnarray}
    {\bf F} &=& F_0 \left[ \sin(k_\textrm{TG} x) \cos(k_\textrm{TG} y) 
                \cos(k_\textrm{TG} z) \hat{x} \right. 
                \nonumber \\
         {} && \left. -\cos(k_\textrm{TG} x) \sin(k_\textrm{TG} y) 
                \cos(k_\textrm{TG} z) \hat{y} \right] ,
\end{eqnarray}
where $F_0$ is the amplitude of the force, which was kept constant in time.
Although the forcing injects energy directly only into the $x$- and
$y$-components of the velocity, the resulting flow is three-dimensional because
of pressure gradients that excite the remaining component of the velocity field.
This forcing injects no energy in the 2D modes, and only affects directly a few
modes in Fourier space corresponding (for the choice $k_\textrm{TG} =1$)
to the mode ${\bf k}=(1,1,1)$ in the first quadrant, and the modes
obtained after reflections across the axes in Fourier space. As will 
become evident later, forcing only these modes is better for the 
excitation of waves than forcing, e.g., all modes in a spherical shell 
in Fourier space. Finally, Taylor-Green forcing is of interest as it 
mimics the flow generated in some experiments using two
counter-rotating disks \cite{Bourgoin02,Ponty05}.

As explained above, the forcing was applied at modes 
such as ${\bf k}=(1,1,1)$, which results in a forced wave number 
$k_\textrm{F} = |{\bf k}| = \sqrt{3}$, and in a forced length scale 
$L = 2\pi/k_\textrm{F} = 2\pi/\sqrt{3}$. The amplitude of the
force was $F_0 = 0.277$ in all the runs, and this value 
was chosen to have an r.m.s.~velocity close to 1 in the
turbulent steady state in the absence of rotation (in practice, 
$U \approx 0.9$ and fluctuates around this value in time). 
The kinematic viscosity was $\nu = 6.5 \times 10^{-4}$, 
resulting in a Reynolds number in the turbulent steady state 
$\textrm{Re} \approx 5000$.

Three runs were done using the following procedure. 
First, a simulation with no rotation ($\Omega = 0$) was done starting 
from the fluid at rest (${\bf u} = 0$), and applying the Taylor-Green 
forcing until the system reached a turbulent steady state. This run 
was continued for $12$ large scale turnover times. Using the 
final state of this run as an initial condition, two other runs were 
done, respectively with $\Omega = 4$ and $8$, and keeping the 
external force and all other parameters the same. Both runs were 
also evolved for $12$ large scale turnover times. This results in 
three runs with Rossby numbers respectively of 
$\textrm{Ro} \approx \infty$, $0.03$, and $0.015$. The last $6$ 
turnover times of each run (in all cases, after the system reached 
the turbulent steady state) were used to compute the spectra and 
correlation functions presented below. For the analysis, data was 
saved with a time cadence $\Delta t = 0.01$.

\begin{figure*}
    \centering
    \includegraphics[width=16cm]{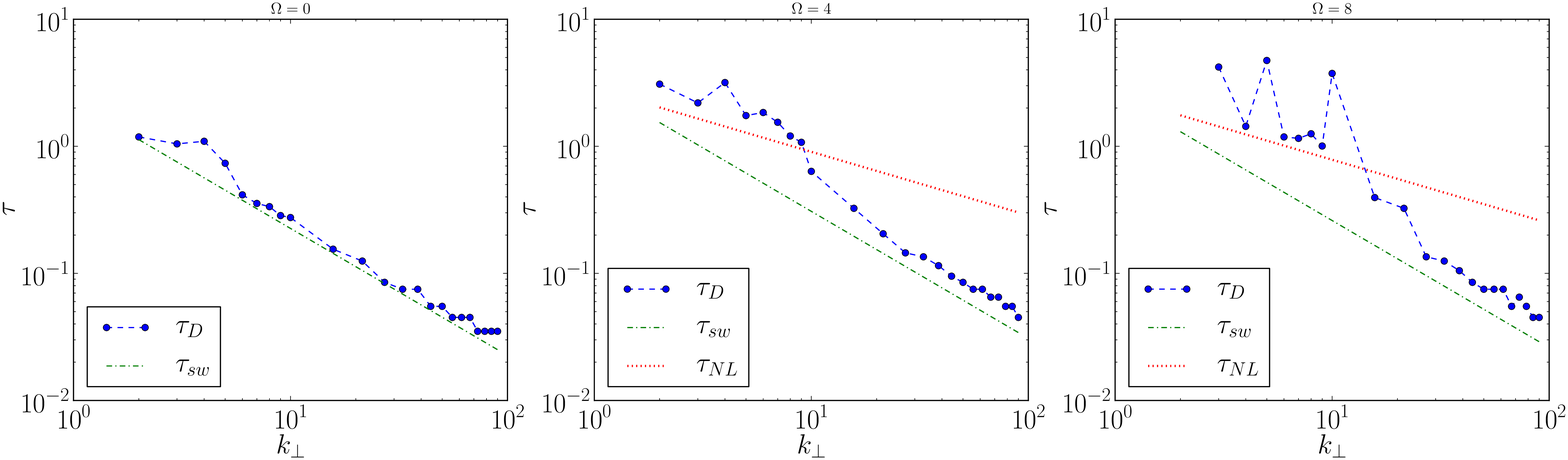} \\
    \includegraphics[width=16cm]{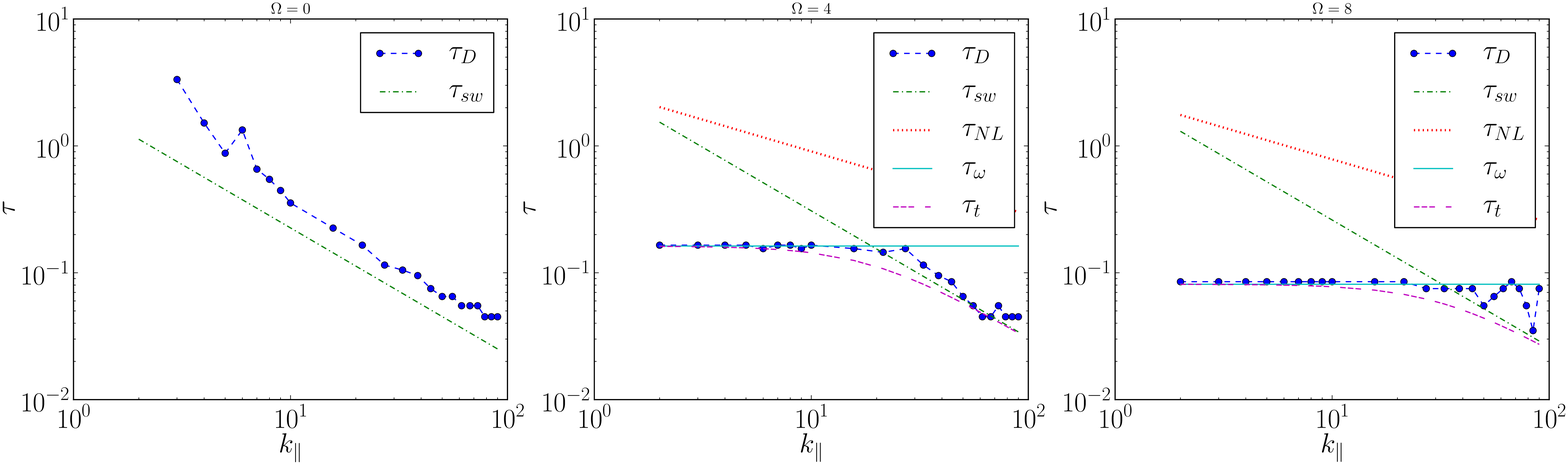}
    \caption{{\it (Color online)} {\it Top:} Decorrelation time $\tau_D$ for the 
        modes with $k_\parallel=0$, as a function of $k_\perp$, and for the 
        three different values of $\Omega$. {\it Bottom:} Same for the modes 
        with $k_\perp=0$, as a function of $k_\parallel$. In all figures, the data 
        corresponds to the dots connected by dashed lines; the wave
        period $\tau_\omega$, the sweeping time $\tau_\textrm{sw}$,
        the nonlinear time $\tau_\textrm{NL}$, and the total effective
        time $\tau_\textrm{t}$ are given as references.}
    \label{fig:tauD_axis}
\end{figure*}

\section{Analysis}

\subsection{\label{sec:behavior}Behavior of the runs in wavenumber space}

Before proceeding to the analysis of the wavenumber and frequency spectrum, and
to the study of the decorrelation time for each mode, we present some spectra
in wavenumber space, as is often done to characterize turbulent flows. Besides
being useful to characterize the runs, these spectra will be also important to
identify the behavior of the different modes depending on what
dynamical times are expected to be dominant.

Figure \ref{fig:spectrum} shows the isotropic energy spectrum 
  $E(k)$ for the run with $\Omega=0$, and the reduced perpendicular energy 
  spectrum $E(k_\perp)$ for the two runs with rotation. The reduced
perpendicular spectrum is obtained by integrating the power spectrum 
of $\hat{\bf u}({\bf k},t)$ over cylindrical shells around the axis of
rotation, and averaging in time to obtain a spectrum that depends 
only on $k_{\perp} = \sqrt{k_x^2+k_y^2}$. In the absence of rotation, 
the isotropic spectrum has a narrow range of wave numbers 
  compatible with Kolmogorov scaling, followed by a bottleneck and 
  a dissipative range. In the rotating case the spectrum becomes 
steeper, as expected.

The axisymmetric energy spectrum $e(k_\perp,k_\parallel)$, obtained after
integrating the power of $\hat{\bf u}({\bf k},t)$ only over the azimuthal
angle in Fourier space, provides more information on the anisotropy of the
flow. As rotation is along the $z$ axis, $k_\parallel = k_z$. Figure 
\ref{fig:2Dspectrum} shows contour plots of
$e(k_\perp,k_\parallel)/\sin(\theta_k)$ for the runs with $\Omega=4$
and with $\Omega=8$, and where 
$\theta_k = \arctan(k_\perp/k_\parallel)$ is the colatitude in Fourier
space. For an isotropic flow ($\Omega = 0$), contours of 
$e(k_\perp,k_\parallel)/\sin(\theta_k)$ are circles. As rotation is
increased, energy becomes more concentrated near the axis with 
$k_\parallel=0$.

\begin{figure*}
    \centering
    \includegraphics[width=16cm]{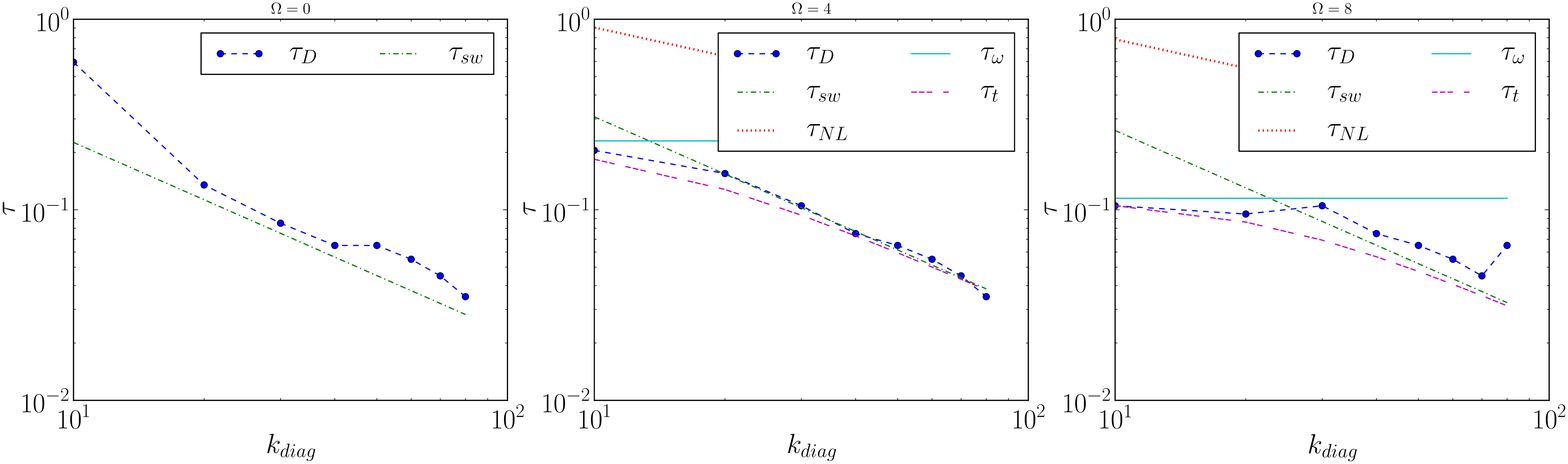}
    \caption{{\it (Color online)} {\it Top:} Decorrelation time $\tau_D$ for 
        the modes with $k_\perp=k_\parallel$, as a function of
        $k=\sqrt{k_\perp^2+k_\parallel^2}$. Labels for the curves are as in
        Fig.~\ref{fig:tauD_axis}.}
    \label{fig:tauD_diag}
\end{figure*}

Based on the previous discussion on wave turbulence theory, and on previous
studies of decorrelation times in isotropic turbulence 
\cite{Chen89,Nelkin90,Sanada92} and in rotating flows \cite{Favier10}, we
can expect several timescales to be relevant for our studies. These timescales
depend on the wave vector, and assuming the shorter one dominates the dynamics,
different regions in the  axisymmetric energy spectrum $e(k_\perp,k_\parallel)$
can be defined. The first timescale is the period of the waves
\begin{equation}
    \tau_\omega ({\bf k}) = C_\omega \frac{k}{2 \Omega k_\parallel} ,
\label{eq:tauomega}
\end{equation}
where $C_\omega$ is a dimensionless constant of order unity.

This time should be compared with the eddy turnover time 
$\tau_\textrm{NL} \sim 1/[k\sqrt{k E(k)}]$. Simple phenomenological
arguments suggest the isotropic energy spectrum in the inertial
range of rotating turbulence follows $E(k) \sim \epsilon^{1/2} \Omega^{1/2}
k^{-2}$ \cite{Zhou95,Muller07,Mininni12}. Then, a possible estimation of the
eddy turnover time is
\begin{equation}
    \tau_\textrm{NL} ({\bf k}) = 
        C_\textrm{NL} \frac{1}{\epsilon^{1/4} \Omega^{1/4} k^{1/2}} ,
\label{eq:tauNL}
\end{equation}
where $C_\textrm{NL}$ is another dimensionless constant of order
unity, and where $\epsilon$ is the energy injection rate. It is worth
noticing that the spectrum of rotating turbulence is actually
anisotropic and dependent on $k_\parallel$ and $k_\perp$ instead of
simply on $k$. However, for the purpose of the discussion here, and as
we are only concerned with order of magnitude estimation of the
timescales, we will use the simplest isotropic expression of $E(k)$.

Sweeping may be the dominant process in the decorrelation of Fourier modes
when the sweeping time becomes shorter than the wave period, as is the case in
isotropic turbulence \cite{Tennekes75,Chen89,Nelkin90,Sanada92}, and
as also found in simulations of rotating turbulence at lower resolution
\cite{Favier10}. The sweeping time is
\begin{equation}
    \tau_\textrm{sw} ({\bf k}) = C_\textrm{sw} \frac{1}{Uk} ,
\label{eq:tausw}
\end{equation}
where $C_\textrm{sw}$ is a dimensionless constant of order
unity. Finally, phenomenological theories of rotating turbulence (see, 
e.g., \cite{Zhou95,Muller07,Mininni12}) often also consider an energy
cascade transfer time 
$\tau_\textrm{tr} \sim \tau_\textrm{NL} (\tau_\textrm{NL}/\tau_\omega)$,
where the ratio of timescales between parenthesis expresses the fact
that waves slow down the energy cascade.

In Fig.~\ref{fig:2Dspectrum} we indicate two curves, corresponding to the 
modes that satisfy the relations 
$\tau_\omega ({\bf k}) = \tau_\textrm{sw} ({\bf k})$, and 
$\tau_\omega ({\bf k}) = \tau_\textrm{NL} ({\bf k})$. Modes inside the
region enclosed by the former curve have the wave period faster than any other
time, and we should expect correlation functions 
$\Gamma_{ii}({\bf k},\tau)$ for these modes to be harmonic. Modes outside
that region should decorrelate with the fastest time, which is the sweeping
time. Finally, modes outside the region enclosed by the latter curve have the
eddy turnover time shorter than the wave period, and as a result those modes
cannot be considered as waves slowly modulated by eddies. In fact, for 
those modes the effect of rotation should be negligible. These 
considerations will be important in the next subsection. To plot the curves 
$\tau_\omega ({\bf k}) = \tau_\textrm{sw} ({\bf k})$ and 
$\tau_\omega ({\bf k}) = \tau_\textrm{NL} ({\bf k})$, we used 
$C_\omega = 1.3$,  $C_\textrm{NL} = 1$, $C_\textrm{sw} = 2.3$; these
values were obtained from the analysis of the data in Sec.~\ref{sub:times}. 

\begin{figure*}
    \centering
    \includegraphics[width=14cm]{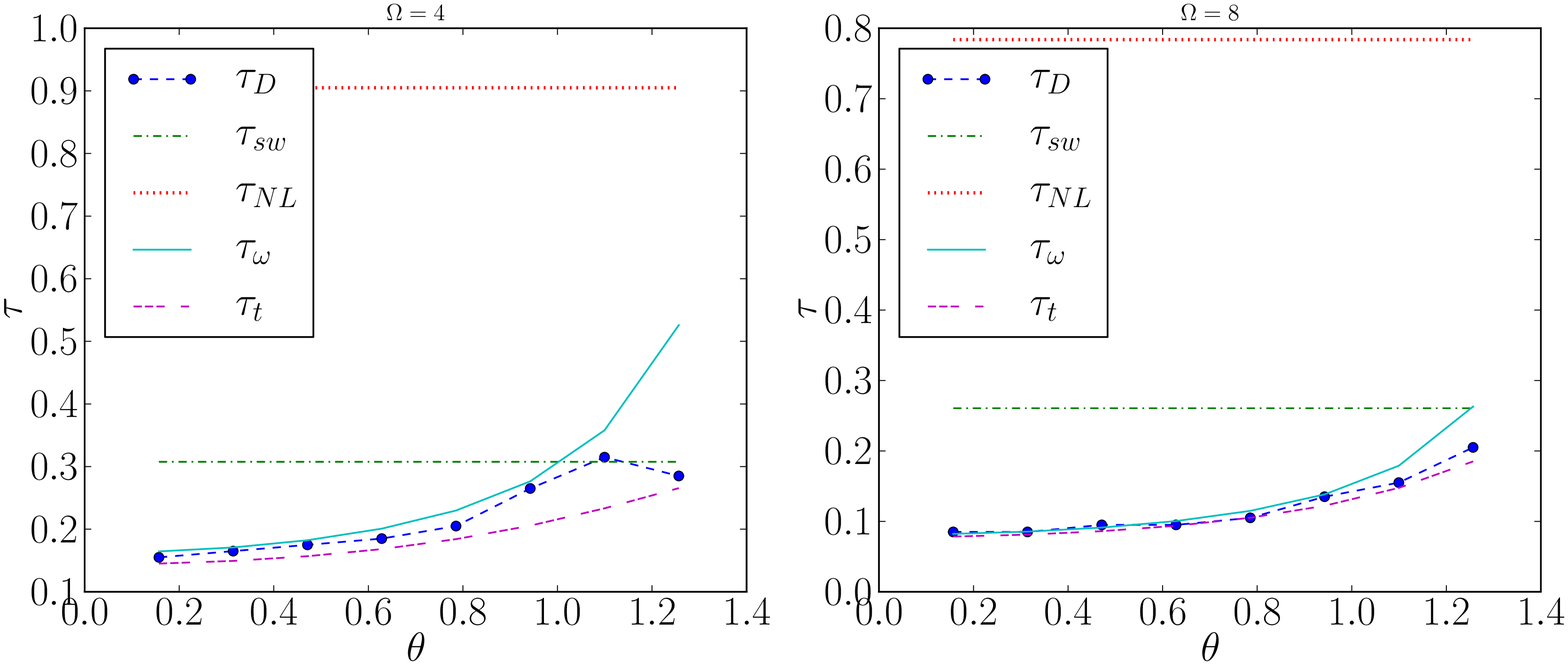} \\
    \includegraphics[width=14cm]{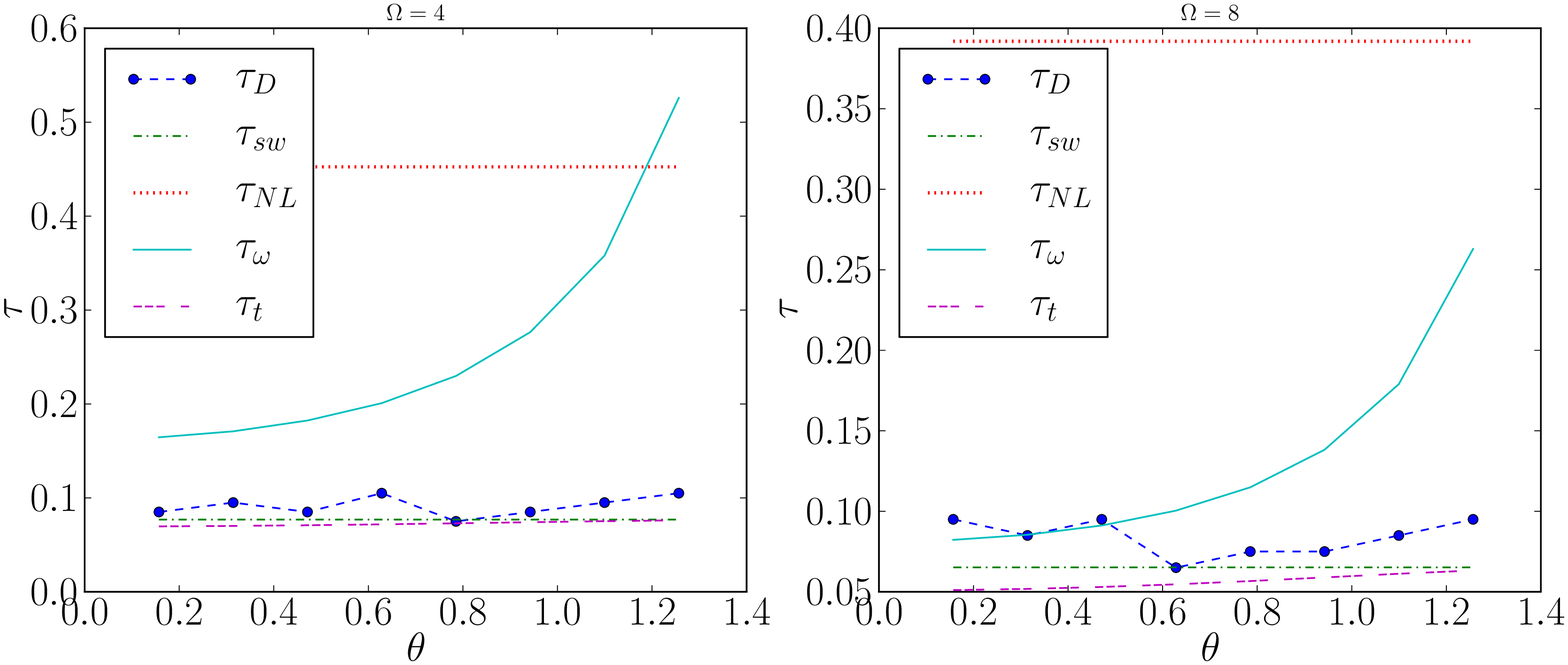}
    \caption{{\it (Color online)} Decorrelation time $\tau_D$ for modes with 
        $k=$constant, as a function of $\theta_k$ (the colatitude in Fourier 
        space). {\it Top:} $\tau_D(\theta_k)$ for $k=10$. {\it Bottom:} Same for 
        $k=40$. Labels for the curves are as in Fig.~\ref{fig:tauD_axis}.
       }
    \label{fig:tauD_angle}
\end{figure*}

In table \ref{table:energies} the fraction of the energy that is contained in 2D
modes, in modes with $\tau_\omega < \tau_\textrm{sw}$, and in modes with
$\tau_\omega < \tau_\textrm{NL}$ is shown for the simulations with $\Omega = 4$
and 8. As mentioned previously, energy becomes more concentrated near the axis
with $k_\parallel=0$ in the presence of rotation.  However, the fraction of the
energy in 2D modes actually decreases as $\Omega$ is increased from 4 to 8, and
more of the injected energy remains in the modes with $\tau_\omega <
\tau_\textrm{sw}$ (although a significant portion of the energy, $\approx 30\%$,
still escapes outside this region and concentrates in the 2D modes). The energy
that is concentrated in these modes comes solely from the leakage from the 3D
modes, as no energy is injected directly into the 2D modes by the forcing we are
using.

The final motivation to use Taylor-Green forcing now becomes apparent. The modes
excited by the forcing are in the region $\tau_\omega < \tau_\textrm{sw}$, and
favors modes dominated by the waves. As a result, all energy in the region with
$\tau_\omega > \tau_\textrm{sw}$, and in the region with $\tau_\omega >
\tau_\textrm{NL}$, can only be accounted for by the nonlinear transfer of energy
from the modes dominated by the wave time. Weak turbulence theories
\cite{Galtier03} cannot account for this transfer.

It should also be noted that in Fig.~\ref{fig:2Dspectrum} the energy does not
accumulate near the modes with $\tau_\omega = \tau_\textrm{NL}$, as it is
expected in theories dealing with the concept of critical balance
\cite{Nazarenko11}. In critical balance, it is argued that in the case of strong
turbulence, energy in the weak turbulence modes cascades towards larger values
of $k_\perp$, while energy in modes with $\tau_\omega < \tau_\textrm{NL}$ (which
are outside the domain of weak turbulence, and are, therefore, strong) cascade
inversely towards smaller values of $k_\perp$ \cite{Nazarenko11,Nazarenko}.
This establishes a balance with $\tau_\omega = \tau_\textrm{NL}$; energy
accumulates in the modes that satisfy this balance and then cascades towards
larger values of $k$ along this curve. No such accumulation is visible in
Fig.~\ref{fig:2Dspectrum}, and as only modes with $\tau_\omega <
\tau_\textrm{sw}$ are forced, the energy in the domain $\tau_\omega >
\tau_\textrm{NL}$ can only come from a transfer from the wave modes to the
vortical modes in the direction opposite to that needed to establish the
balance.

\subsection{Wave vector and frequency spectrum}

Figure \ref{fig:wspectra} shows the wave vector and frequency spectrum
$E_{11}({\bf k}, \omega)/E_{11}({\bf k})$ for different values of ${\bf k}$,
where
\begin{equation}
    E_{11} (\mathbf{k}) = \int E_{11}({\bf k}, \omega) \, \text{d} \omega .
\end{equation}
With this choice for the normalization, the frequencies that
concentrate most of the energy for each ${\bf k}$ are more clearly visible.

When $k_x=0$, $k_y=0$, and $k_z$ is varied, most of the energy is concentrated
near $\omega = 2 \Omega$, especially for $k_z<10$. For larger values of
$k_z$ the width of the band that concentrates most of the energy increases
(compare this with the regions in Fig.~\ref{fig:2Dspectrum} corresponding to
modes with $\tau_\omega ({\bf k}) < \tau_\textrm{sw}  ({\bf k})$, and to modes
with $\tau_\omega ({\bf k}) < \tau_\textrm{NL}$). The wave vector and frequency
spectrum for the other components of the velocity were also calculated, showing
similar behavior.

When $k_x=0$, $k_y$ is kept fixed, and $k_z$ is varied, most of the energy is
still concentrated near the linear dispersion relation $\omega = 2 \Omega
k_z/\sqrt{k_y^2+k_z^2}$ in the cases with $k_y=1$ and $k_y=5$. However, for
$k_y=10$ and larger, energy is more evenly distributed among all values of
$\omega$, and most of the energy is in the modes with $\omega \approx 0$.

Leaving aside the energy in the modes with $\omega \approx 0$, the modes that
concentrate energy near the linear dispersion relation could in principle be
treated by weak turbulence theories, where energy is transferred through wave
interactions. But it is worth pointing out also that the accumulation of energy
in these fast modes makes (at least for a subset of the wave numbers) some
magnitudes in the turbulent flow treatable by RDT
\cite{Cambon89,cambon_linear_1999}. This has been used to study the early time
evolution of the system when rotation is turned on in an initially isotropic
flow \cite{brethouwer_effect_2005}).

\subsection{\label{sub:times}Correlation functions and decorrelation times}

Figure \ref{fig:correlation} shows the time correlation function 
$\Gamma_{11}({\bf k}, \tau/\tau_t)$ for the $x$-component of the velocity, 
with the time being normalized by a total effective time $\tau_t$, and
for different modes in Fourier space. The total effective time is
defined as
\begin{equation}
    \left(\frac{1}{\tau_\textrm{t}} \right)^2 = 
        \left(\frac{1}{\tau_\omega} \right)^2 
        + \left(\frac{1}{\tau_\textrm{sw}} \right)^2,
\label{eq:taut}
\end{equation}
such that for $\tau_\omega \ll \tau_\textrm{sw}$, 
$\tau_\textrm{t} \approx \tau_\omega$, and for 
$\tau_\omega \gg \tau_\textrm{sw}$, 
$\tau_\textrm{t} \approx \tau_\textrm{sw}$. With this definition, 
$\Gamma_{11}(\tau/\tau_t) = 1/e$ at a time $\tau/\tau_t$ of order
unity for all modes.

By inspection of these functions, we identified three different
behaviors that are illustrated by a few modes in the figure. Modes near the
$k_\parallel$ axis (and for sufficiently small $k_\parallel$) have $\Gamma \sim
\cos(\omega_{\bf k} \tau)$, the behavior expected for waves ($\tau_t \approx
\tau_\omega$ for these wavenumbers). As $k_\parallel$ is increased (and as
$k_\perp$ is increased as well), the correlation functions still show a
wave-like behavior, but also display a slower decay in their modulation in a 
time that can be associated with the eddy timescale $\tau_\textrm{NL}$ 
(following Eq.~\ref{eddiewaves}, the eddy or ``slow'' timescale $T$ is of
order one when $t\sim {Ro}^{-1}$; in that timescale the decay time of the
correlation functions is given by $\tau_\textrm{NL}$). It is interesting that 
the slow modulation takes place on a timescale of the order of ${Ro}^{-1}$, 
as some studies suggest that this is the time in which energy is transfered 
outside the region of weak turbulence in Fourier space, and for which 
rotating turbulence becomes strong \cite{Chen05}. Finally, for even larger 
values of the wavenumber, the correlation functions decay exponentially 
and resemble those obtained in the absence of rotation. Correlation 
functions for the other components of the velocity were also calculated, 
showing similar behavior.

From the correlation functions the decorrelation time $\tau_D$ can be
measured directly. Figure \ref{fig:tauD_axis} shows $\tau_D$ for the 
$x$-component of the velocity for modes along the $k_y$ axis (i.e., 
for $k_x=k_z=0$), and along the $k_z$ axis (i.e., for $k_x=k_y=0$). In 
the case with $k_\parallel = 0$, the modes seem to decorrelate with
the sweeping time independently of the value of $\Omega$, as also
found in \cite{Favier10}, although in the rotating case and for small 
$k_\perp$ ($k_\perp \lesssim 10$) the behavior seems to be compatible 
with $\tau_D \sim \tau_\textrm{NL}$. Note that in the runs with
rotation, these modes correspond to the 2D or ``slow'' modes, with 
zero wave frequency. As a result, these modes can only be vortical.

The modes with $k_\perp = 0$ decorrelate with the sweeping time in the run
without rotation, but in the runs with rotation $\tau_D$ has a transition at
$\tau_\omega({\bf k}) = \tau_\textrm{sw}({\bf k})$. For $k_\parallel$ such that
$\tau_\omega({\bf k}) < \tau_\textrm{sw}({\bf k})$, $\tau_D \approx
\tau_\omega$. For $k_\parallel$ such that $\tau_\omega({\bf k}) >
\tau_\textrm{sw}({\bf k})$, $\tau_D \approx \tau_\textrm{sw}$. This
transition was also found in the simulations in \cite{Favier10}. The nonlinear
time plays no clear role in the decorrelation. A similar behavior is obtained if
${\bf k}$ is varied along the diagonal with $k_y=k_z$ and $k_x=0$ (see
Fig.~\ref{fig:tauD_diag}), or if $k$ is kept fixed and the colatitude $\theta_k$ 
in Fourier space is varied (see Fig.~\ref{fig:tauD_angle}). This former case is
interesting as for small values of $k$ the decorrelation time $\tau_D$ is closer
to the wave period, while for larger values of $k$ the decorrelation is closer
to the sweeping time (i.e., approximately constant with
$\theta_k$). As explained above, the modes dominated by the
  faster timescale $\tau_\omega$ satisfy conditions akin to the
  hypothesis made in RDT \cite{Cambon89,cambon_linear_1999}.

The dimensionless constants $C_\omega$, $C_\textrm{NL}$, and $C_\textrm{sw}$ 
in Eqs.~\eqref{eq:tauomega}, \eqref{eq:tauNL}, and \eqref{eq:tausw} were 
chosen from the data in Figs.~\ref{fig:tauD_axis}, \ref{fig:tauD_diag}, and 
\ref{fig:tauD_angle}, and are the same for all runs (independently of the mode 
${\bf k}$ studied, and of the value of $\Omega$). Note these amplitudes only 
account for an arbitrariness in the definition of the decorrelation time, which
from the time correlation function and as explained above can be defined 
based on the time to decay to $1/e$ of its value, based on its half-width, 
or on an integral timescale. 

From these observations, it becomes apparent that the effective time 
$\tau_t$ gives a good approximation to the actual decorrelation time 
$\tau_D$ in all figures. It is interesting that the choice to average the relevant 
times in Eq.~\eqref{eq:taut} is similar to the choice used in the simplest EDQNM 
models of rotating turbulence to estimate the eddy damping 
\cite{Cambon89,Cambon97}.

From these figures, we can also conclude that the modes in the region enclosed
by the curve $\tau_\omega = \tau_\textrm{sw}$ in Fig.~\ref{fig:2Dspectrum} have
wave-like behavior with decorrelation dominated by the waves, while the rest of
the modes are dominated by sweeping effects (similar to what happens in
isotropic and homogeneous turbulence \cite{Chen89}).

\subsection{\label{sub:anisotropy}Decorrelation times and anisotropy}

The fact that modes with $\tau_\omega > \tau_\textrm{sw}$ have
decorrelation dominated by the sweeping time should not be interpreted
as that the effects of waves and of rotation are negligible for these 
modes. This is evidenced quite clearly in Fig.~\ref{fig:2Dspectrum}, where 
anisotropic spectral distribution of energy can be observed even for modes 
with $\tau_\omega > \tau_\textrm{sw}$. Isotropy is expected to be
recovered at the Zeman wavenumber $k_\Omega$ for which 
$\tau_\omega(k_\Omega) = \tau_\textrm{NL}(k_\Omega)$
\cite{Zeman1994}. This wavenumber is equivalent to the Ozmidov
wavenumber in a stratified flow, and that isotropy is recovered in 
rotating turbulence at that wave number has been recently confirmed 
in high resolution numerical simulations \cite{Mininni12}.

From the expressions in Sec.~\ref{sec:behavior}, assuming that 
when isotropy is recovered $k_\perp \approx k_\parallel$ and therefore 
$k \approx \sqrt{2} k_\parallel$, we can write the condition 
$\tau_\omega = \tau_\textrm{NL}$ at $k=k_\Omega$ as
\begin{align*}
        \frac{C_\textrm{NL}}{\epsilon^{1/4} \Omega^{1/4} k^{1/2}_\Omega} 
        &= \frac{C_\omega}{\sqrt{2} \Omega}
        \\
        \Rightarrow
        k_\Omega &= C_\Omega \left( \frac{\Omega^3}{\epsilon}
        \right)^{1/2} ,
\end{align*}
where $C_\Omega = 2 (C_\textrm{NL}/C_\omega)^2 \approx 1.18$.

This expression is compatible with the one found in 
\cite{Mininni12}, where $C_\Omega = 1$ was found from direct 
observation of the scale at which isotropy was recovered. For the 
simulation with $\Omega = 4$, $k_\Omega \approx 150$ (which lies 
in the dissipative range of the simulation), and for the simulation 
with $\Omega = 8$, $k_\Omega \approx 460$ (which lies outside the 
domain of resolved scales).

It is interesting that the characteristic timescales 
discussed here present another interpretation of the Zeman scale: 
isotropy is recovered not when all modes satisfy the condition 
$\tau_\textrm{NL} \le \tau_\omega$ (which happens in the runs 
with $\Omega = 4$ and $8$ at much larger wave numbers, see 
Fig.~\ref{fig:2Dspectrum}), but when a significant fraction of the 
modes satisfy this condition (i.e., when the modes in the diagonal 
with $k_\perp \approx k_\parallel$ satisfy the equality of 
timescales).

\section{Conclusions}

\begin{figure}
    \centering
    \includegraphics[width=8cm]{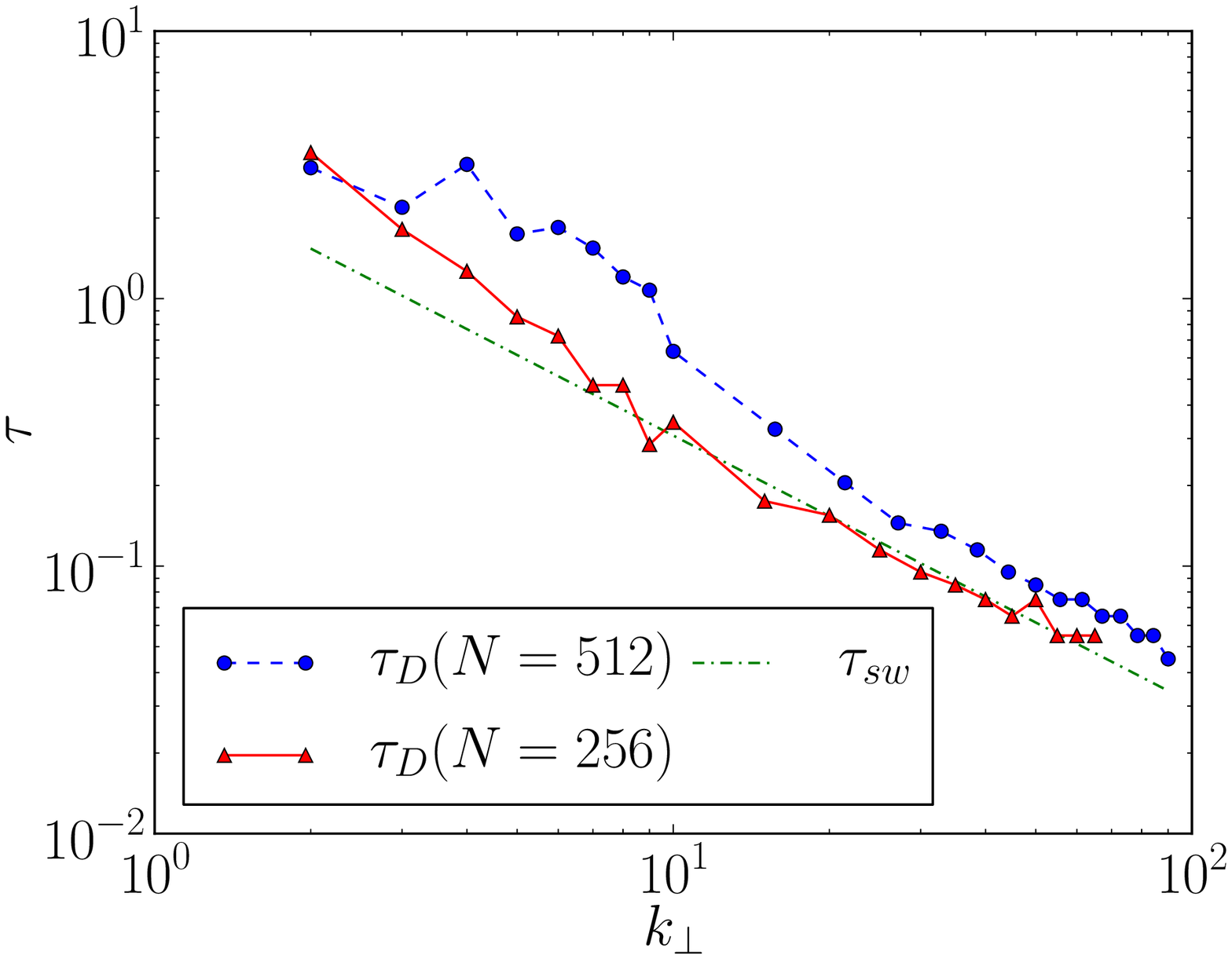} \\
    \includegraphics[width=8cm]{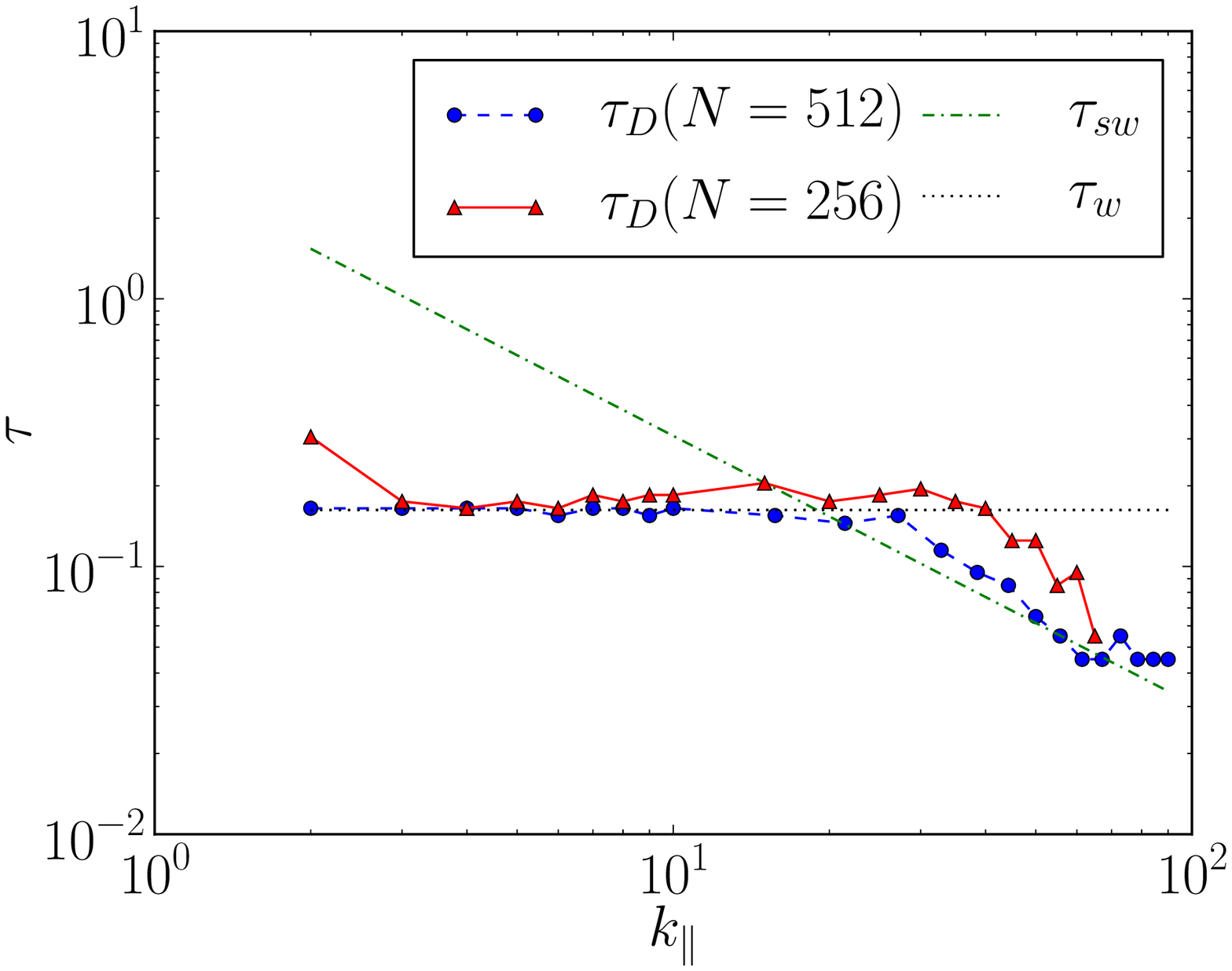}
    \caption{{\it (Color online)} {\it Top:} Decorrelation time $\tau_D$ for the 
        modes with $k_\parallel=0$ as a function of $k_\perp$, for runs with
        same value of $\Omega$ but different spatial resolution $N$ and
        Reynolds number. The time $\tau_D (N=512)$ is for the run with 
        $N=512$ linear resolution, while $\tau_D (N=256)$ is for the run 
        with $N=256$. {\it Bottom:} Same for the modes with $k_\perp=0$, 
        as a function of $k_\parallel$. In all figures, the data
        corresponds to the symbols connected by dashed lines. The
        curves scaling with wavenumber as the wave
        period $\tau_\omega$, the sweeping time $\tau_\textrm{sw}$,
        and the total effective time $\tau_\textrm{t}$ are shown as 
        references.}
    \label{Rfig:tauD_axis}
\end{figure}

The results presented here indicate that: (1) A significant fraction of the
energy is concentrated in modes with $\omega \approx 0$, which have correlation
functions corresponding to that of strong turbulence (i.e., vortical modes). In
this respect, the simulations are limited to finite domains and we cannot
conclude from this what the behavior is for a homogeneous (infinite) flow as is
often considered in wave turbulence theories. (2) For modes with $\tau_\omega
\ll \tau_\textrm{sw}$, a significant fraction of the remaining energy is
concentrated near the dispersion relation $\omega({\bf k})$ of inertial waves.
However, the dispersion relation is not visible anymore in $E({\bf k}, \omega)$
as the wave number is increased. (3) For $\tau_\omega < \tau_\textrm{sw}$, the
correlation functions behave as expected for modes in weak turbulence theories:
waves slowly modulated by eddies. For $\tau_\omega \ll \tau_\textrm{sw}$, waves
dominate and the correlation functions are harmonic in $\tau$. The
rest of the modes decorrelate with the sweeping time. However,
the effects of rotation only become negligible (e.g., to reobtain
isotropy) for modes with $\tau_\textrm{NL} \le \tau_\omega$.

Previous studies of the time correlation function at lower
Reynolds number and at lower spatial resolution, using $256^3$ grid
points, obtanied similar results \cite{Favier10}. This indicates
that the results are not very sensitive to the value of the Reynolds 
number, at least in the range of parameters considered in the
present study. To further confirm this, in the Appendix we present a 
comparison of our results with simulations with $256^3$ grid points 
and with $\textrm{Re} \approx 3100$. Studying flows with higher 
Reynolds numbers is currently out of our reach, as computation of time
correlation functions and of the wave number and frequency spectrum 
require storage of vast amounts of data which increase rapidly as 
spatial resolution is increased. Such higher Reynolds numbers
studies would determine the degree of applicability of the current 
results to the fully developed turbulence regime.

Finally, the results presented here were obtained for a forcing
function that only excites a few modes in the region of Fourier space 
dominated by the wave timescale. For isotropic, random, 
delta-correlated in time forcing, preliminary studies indicate that 
at the same Rossby number the effect of the waves is weaker.

\begin{acknowledgments}
PCdL, PJC, PDM and PD acknowledge support from grants No.~PIP
11220090100825, UBACYT 20020110200359, and PICT 2011-1529 
and 2011-1626.
\end{acknowledgments}

\appendix*
\section{Simulations with lower Reynolds number} 

We briefly compare here the results obtained from the simulations with
$512^3$ grid points and $\textrm{Re} \approx 5000$, with simulations 
at lower resolution and lower Reynolds number. We performed a set of 
simulations with the same Rossby numbers as the $512^3$ simulations, 
but with larger viscosity and using $N^3=256^3$ grid points, resulting 
in a Reynolds number $\textrm{Re} \approx 3100$.

Figure \ref{Rfig:tauD_axis} presents a comparison of the decorrelation 
times in the simulations with $256^3$ and $512^3$ grid points, both
with $\Omega=4$. This figure is equivalent to
Fig.~\ref{fig:tauD_axis}, in which only the results for the $512^3$
simulation were shown. The sweeping time and the wave period are the 
same for both simulations, as the forcing amplitude and the value of 
$\Omega$ were kept the same. No significative differences are observed 
between both simulations, and the decorrelation time presents the same 
behavior with $k_\parallel$ and $k_\perp$ in both cases. Small differences 
observed are due to the fact that the r.m.s.~velocity is not exactly 
the same in both runs.

We also verified that other quantities (such as $\Gamma_{11}$,
the correlation functions) also act similarly in the $\textrm{Re}=3100$
and $\textrm{Re}=5000$ cases, lending some confidence that the results
described above are not very sensitive to the Reynolds number
at the modest values considered here.

%

\end{document}